# Giant Electro-Viscous Effects in Polar Fluids with Paraelectric–Modulated Antiferroelectric–Ferroelectric Phase Sequence

*Hiroya Nishikawa[1], Péter Salamon,[2,*] Marcell Tibor Máthé,[2] Antal Jákli,[3,4,*] and Fumito Araoka[1,*]*

1 RIKEN Center for Emergent Matter Science (CEMS), 2-1 Hirosawa, Wako, Saitama 351-0198, Japan

2 HUN-REN Wigner Research Centre for Physics, P.O. Box 49, Budapest H-1525, Hungary

3 Material Science Graduate Program and Advanced Materials and Liquid Crystal Institute, Kent State University, Kent, Ohio 44242, USA

4 Department of Physics, Kent State University, Kent, Ohio 44242, USA

*: Authors for correspondence: fumito.araoka@riken.jp, ajakli@kent.edu, salamon.peter@wigner.hun-ren.hu

**Abstract**

The recently discovered ferroelectric nematic liquid-crystal material DIO exhibits an antiferroelectric (AF) phase, characterized by a sinusoidally modulated structure between the paraelectric (P) and ferroelectric (F) nematic phases. Although these sinusoidal modulated structures associated with the P–AF–F phase sequence is commonly observed in solid ferroelectrics, their presence in soft matter systems is scarce. This study is aimed at examining the macroscopic properties of DIO, identifying unique rheological properties, such as switching between shear thinning and shear thickening behaviors at certain shear rate in the ferroelectric nematic phase. Additionally, a significant electroviscous effect is observed, with the viscosity increasing by 70 times under an ultra-low electric field of 0.15 V $\mu m^{-1}$ at the AF–F transition.

1. Introduction

A small fraction of crystals with three-dimensional (3D) long-range order exhibits a unique incommensurate (IC) structure, which lacks translational periodicity in at least one direction [1]. In IC materials, internal stress may induce a periodic modulation where the ratio of the modulation period to the fundamental lattice one is an irrational number. If this ratio is rational, the structure is classified as commensurate (C). IC structures often emerge through phase transitions of normal (N)–incommensurate (IC)–commensurate (C) structures phase transition sequence. This phenomenon occurs in a wide range of materials, including dielectrics [1,2], low-dimensional conductors [3], magnetic materials (magnetic helical structures) [4], surface structures [5], intercalation compounds [6], nonstoichiometric compounds [7], liquid crystals (LCs) [8], composite crystals [9], and superconducting oxides [10]. Among these, dielectrics, especially ferroelectrics, are the dominant IC materials, with typical examples including sodium nitrite ($NaNO_2$) [11] and thiourea ($SC(NH_2)_2$) [12]. These molecular crystals, composed of molecules with permanent dipoles, undergo spontaneous polarization owing to molecular rotation, resulting in the N–IC–C phase transition. In these cases, N, IC, and C correspond to the paraelectric (P), sinusoidally modulated antiferroelectric (AF), and ferroelectric (F) phases, respectively.

IC structures are observed not only in solid materials but also in soft matter, such as LCs. For example, the improper ferroelectric chiral smectic (SmCα*) phase exhibits an IC nanoscale helical pitch [8]. In this case, the phase sequence is SmA (paraelectric, P)–SmCα* (ferrielectric, FI)–SmC$_A$* (antiferroelectric, AF), i.e., P–FI–AF. Recently, ferroelectric nematic (N$_F$) LCs have garnered significant research interest owing to their excellent polarization properties (i.e., high polarization density and nonlinear optical coefficient) [13] and other unique effects, such as splay elastic softening [14], novel topology [15], thermo-rotation [16], self-propelled droplets [17], polar fiber formation [18], surface instability [19], superscreening [20], and helicity [21]. One of the archetypal N$_F$ LCs, i.e., DIO [13a] exhibits the M1–M2–MP phase sequence upon cooling, which has been identified as a paraelectric nematic (N)–antiferroelectric smectic Z (SmZ$_A$)–ferroelectric nematic (N$_F$) sequence (Figure 1) [22]. This P–AF–F sequence is similar to that observed in NaNO$_2$ [23] and SC(NH$_2$)$_2$ [24], as well as in recently predicted electric-field-induced phase diagrams [22]. In the N phase of DIO, the material exhibits truly 3D fluidity (pure orientational order without positional order). In contrast, in the N$_F$ phase, the inversion symmetry of the N director (**n**) is disrupted (i.e., **n** ≠ −**n**), resulting in the formation of uniform domains with macroscopic polarization, whose direction can be flipped by suitable electric fields [13e]. In addition to the SmZ$_A$ structure of the AF phase, various models for modulated antiferroelectric structures have been proposed. For instance, Mertelj et al. proposed a splay nematic (Ns) structure [14a] with a polar-modulated nematic director field, based on splay deformation observed in another archetypal N$_F$ LC, i.e., RM734 [13b]. Both SmZ$_A$ and Ns models (referred to as antiferroelectric mesophases (M$_{AF}$) herein), are similar in that the nematic director (**n**) and polarization (**P**) vectors are oriented in the same direction within several nanometers thick local-polar slabs, with alternating polarization directions in adjacent slabs, resulting in an antiferroelectric structure [22,25]. Nacke et al. proposed a model in which the distribution of dipole moments with opposite signs increases continuously along the wave vector (⊥**n**) while traversing slabs, until the polarization direction is completely flipped [26]. This suggests that the M$_{AF}$ phase exhibits a sinusoidally modulated antiferroelectric structure, similar to the IC phase observed in the NaNO$_2$/SC(NH$_2$)$_2$ system.

The appearance of antiferroelectric phases with sinusoidally modulated structures in soft matter systems is extremely rare, and the macroscopic properties associated with phase transitions remain understudied. Here we report rheological studies of DIO and find unique properties, such as switching between shear thinning and shear thickening behavior at certain shear rate in the ferroelectric nematic phase, and an order of magnitude larger electro-viscous effect in DIO with P–AF–F sequence than for RM734 with P–F transition.

## 2. Results and Discussion

In LCs, the director rotation and flow velocity gradient are coupled, and the flow viscosity depends on the director (**n**) orientation. For nematic LCs, Miesowicz's viscosity coefficients $\eta_b$ ($\eta_c$) describe the viscosity when the director is parallel (perpendicular) to the flow velocity and perpendicular (parallel) to the velocity gradient: $\eta_b : v \parallel n, \nabla v \perp n$; $\eta_c : v \perp n, \nabla v \parallel n$. $\eta_b$ and $\eta_c$ correspond to the lowest and highest shear viscosity, respectively [27]. Additionally, an intermediate viscosity, $\eta_a$, arises when the director is perpendicular to the shear plane: $\eta_a: v \perp n, \nabla v \perp n$.] Films of rod-like nematic LCs exhibit shear-thinning behavior [28], with the extent of thinning depending on the film thickness. In general, the relationship between the shear stress and shear rate can be described using the power-law model [29]:

$$\sigma = \kappa(\dot{\gamma})^\alpha, \qquad (1)$$

where $\sigma$, $\kappa$, $\dot{\gamma}$, and $\alpha$ denote the shear stress, flow consistency index, shear rate, and flow behavior index, respectively. For Newtonian fluids, $n = 1$, whereas shear-thinning and shear-thickening behaviors are observed when $n < 1$ and $n > 1$, respectively. Viscosity is defined as $\eta = \sigma/\dot{\gamma}$. Figure 2a displays the dependence of shear stress ($\sigma$) on shear rate in the N, $M_{AF}$, and $N_F$ phases, measured under steady-shear conditions. In the N phase, the shear stress is proportional to the shear rate (above 10–100 s$^{-1}$), indicating Newtonian fluid behavior (Figure 2b). Similarly, Figure 3a shows that the Newtonian behavior is true only for shear rate above 10–100 s$^{-1}$. Below these rates, shear thinning occurs. This is because of the transition from flow-induced to surface alignment, which is characterized by a thickness parameter $e_1 = \sqrt{\frac{K}{\eta \dot{\gamma}}}$, which decreases at increasing shear rate [30]. In contrast, the $N_F$ phase exhibits different slopes at the crossover point ($\dot{\gamma} = 20$ s$^{-1}$). Thus, $n = 0.97 < 1$ and $n = 1.02 > 1$ in Regimes I and II, representing shear-thinning and shear-thickening behaviors, respectively. As shown in Figure 2d, the $M_{AF}$ phase exhibits $\sigma \propto \dot{\gamma}^{0.90}$ and $\sigma \propto \dot{\gamma}^{0.95}$ in Regimes I and II, respectively, representing shear thinning. Figure 3 shows the viscosity as a function of the shear rate in the N and $N_F$ phases. In the N phase, at 95 °C and $\dot{\gamma} = 5$ s$^{-1}$, the viscosity is $\eta = 12 \, mPa \cdot s$ (Figure 3a). At increasing shear rates between 100 and 5000 s$^{-1}$, the viscosity decreases to $\eta = 10.8 \, mPa \cdot s$. Near the N–$M_{AF}$ phase transition, $\eta$ increases to 39 mPa·s at low shear rates, with steady-state conditions attained at high shear rates. Polarized optical microscopy (POM) images (Figure 3b) confirm that the director reorients along the flow direction at higher shear rate. The grainy texture observed at low $\dot{\gamma}$ turn to homogeneous over few hundred s$^{-1}$ indicating the alignment by the shear. Similar to the N phase, shear-thinning behavior is also observed at low shear rates ($\dot{\gamma} < 20\ s^{-1}$) in N$_F$ phase (Figure 3c and 3d). Given that the POM texture changes in the same shear rate region (Figure 3e), the director flow begins at higher shear rate than that in the N phase, likely owing to the higher viscosity (59 mPa·s at $T = 54\ °C$) than that in the N phase. Notably, the viscosity increases in the range $20\ s^{-1} \leq \dot{\gamma}$ reaching a nearly constant value of ~60 mPa·s at higher shear rates. In this regime, a uniform texture is observed (Figure 3e). The Newtonian behavior at high shear rates in the nematic phase likely corresponds to the negligible transient alignment near the surface and the resulting uniform alignment. The transient increase in viscosity in the N$_F$ phase above 20 s$^{-1}$ is likely related to the director realignment and generation of polar defects, a distinct characteristic of the N$_F$ phase. In the N and N$_F$ phases, at high shear rates, the texture appears dark (Figures 3b and 3e), with the polarizer axis parallel to the flow direction. This indicates that the flow orients the director to be in the shear plane, which is characteristic of flow-aligning behavior in nematic LCs [31].

In the M$_{AF}$ phase, the viscosity exceeds 100 mPa·s, which is six and two times that in the N and N$_F$ phases, respectively (Figure 4a). As shown in Figure 2d, 4a, and 4b, the M$_{AF}$ phase exhibits two types of shear-thinning behavior r: regime I and II with apparently distinct shear rate dependence of the viscosity. This property may be attributable to tumbling motion, rather than the shear alignment observed in the N and N$_F$ phases, typically found in smectic LCs [32] (Figure 4c). The temperature dependence of apparent viscosity under constant shear rate and stress was evaluated. As shown in Figure 4d, in the N and N$_F$ phases, the apparent viscosity increases with decreasing temperature, independent of the shear rate. In contrast, in the M$_{AF}$ region, the viscosity is higher at low shear rate of $\dot{\gamma} = 2.59\ s^{-1}$ but decreases below the values observed for the N and N$_F$ phases at high shear rates ($\dot{\gamma} = 5000\ s^{-1}$). This difference also arises under different stress conditions, where the viscosity increases at stresses $\tau < 0.75\ Pa$ and decreasing when $\tau > 2\ Pa$ (Figure 4e).

These results show that a distinct structure exists in the M$_{AF}$ phase, as discussed in the introduction. This could correspond either to an antiferroelectric lamellar order with a periodicity of ~17.5 nm, including 8.8-nm-thick layer pairs with opposite ferroelectric polarization, or a modulated splay-nematic phase with ~17.5 nm periodicity [22,25]. Under confinement, the $M_{AF}$ phase can adopt two geometries, wherein the layers align either normal ("bookshelf") or parallel to the plates. At low shear rates, where the smectic structure is not completely destroyed, the system likely selects an optimal structure for flow during shearing. For example, if the parallel geometry of $M_{AF}$ phase was adopted, each lamellar layer may glide along the shear direction (Figure 4f). However, this is not necessarily the optimal geometry, as the two boundary plates are rarely perfectly parallel. Therefore, the layers near the substrates cannot glide and must adjust their thickness, which is energy-intensive. In contrast, in the bookshelf alignment, the boundary of the layers normal to the substrates may exhibit any shape (Figure 4g). Consequently, the parallel alignment is associated with higher viscosity, and the bookshelf alignment corresponds to lower viscosity. Based on these findings, we propose a mechanism for the dual shear-thinning behavior in Regimes I and II for the $M_{AF}$ phase. The shear-thinning behavior in Regime I reflects the deformation of the lamellar layer in the parallel geometry, which corresponds to high viscosity. Subsequently, the geometry reorients to the bookshelf alignment with lower viscosity, reflecting the additional shear-thinning behavior in Regime II. At high shear rates and stress in the $M_{AF}$ phase, the rheology exhibits two key features. First, in contrast to the N and $N_F$ phases, the optical texture under crossed polarizers (Figure 4c) remains bright, indicating the presence of a significant director component perpendicular to the shear plane, characteristic of tumbling nematics [31]. Second, the $M_{AF}$ phase demonstrates lower effective viscosity relative to the N and $N_F$ phases at extremely high shear rates (Figures 4d and 4e). Similar results have been obtained in the nematic phase of 8CB, where viscosity decreases as the system transitions from flow-aligning to tumbling at high shear rates [33].

Next, we investigated the effect of the *E*-field on viscosity across the three phases. The effect of flow and electric field are conflicting with each other. Thus, when the shear rate is sufficiently small, the vertical alignment of the molecules (with positive dielectric anisotropy) can be realized under the *E*-field, thereby increasing the apparent viscosity. In ferroelectric LCs, the induction of flow or deformation can be easily tuned owing to the coupling between spontaneous polarization and the *E*-field (i.e., $P \times E$). As shown in Figure S1a, in the N phase, the apparent viscosity under the $E_{DC}$-field is smaller than that under the $E_{AC}$-field. This difference is attributable to DC ionic screening, which is negligible in $E_{AC}$ fields. In contrast, in the $M_{AF}$ and $N_F$ phases, the application of $E_{DC}$ is more effective in reorienting the director (Figures S1b and S1c), attributable to the polar interaction between the electric field and polarization of these phases. Figures 5a and 5c show the dependence of apparent viscosity on $E_{AC}$- and $E_{DC}$-fields under constant shear stress ($\sigma$ = 20 Pa) in the N/$M_{AF}$ and $M_{AF}$/$N_F$ regimes, respectively. When $E = 0.15\ V\ \mu m^{-1}$ is applied, the viscosity in the N phase doubles (from 10 to 20 mPa·s at 90 °C), whereas in the $N_F$ phase, it increases eightfold (from 30 to 240 mPa·s at 64 °C). In the $M_{AF}$ phase, although $E = 0.15\ V\ \mu m^{-1}$ is inadequate to saturate the viscosity, the viscosity dramatically increases near the $M_{AF}$–$N_F$ phase transition. Specifically, at 70 °C, the viscosity increases 70-fold, from 18.6 mPa·s to 1.29 Pa·s. Figure 6a shows the zero-field viscosity and $E$-field-induced viscosity as functions of temperature under $\dot{\gamma} = 100\ s^{-1}$. The kink points of viscosity change at the N–$M_{AF}$ phase transition appear at the same temperature, regardless of $E$-field application. In contrast, at the $M_{AF}$–$N_F$ phase transition, the kink point temperature under the $E$-field is 2 °C higher than that under the zero field. This suggests that the $M_{AF}$ phase transforms to the $N_F$ phase upon the application of the $E$-field, i.e., the sufficiently large $E$-field induces phase transition from an antiferroelectric state to a ferroelectric state. The induction of the $N_F$ phase from the isotropic phase upon DC electric field application has been observed for other compounds as well [34].

Next, we examined viscosity changes at the N–$M_{AF}$ and $M_{AF}$–$N_F$ phase transitions under $E$-field. The relative percentage enhancement of viscosity as a percentage is defined as

$$\eta_{\%} = \frac{\eta(E=E_s)-\eta_0}{\eta_0} \times 100, \qquad (2)$$

where $\eta_{\%}$, $\eta(E=E_s)$, and $\eta_0$ denote the percentage increase in saturated viscosity and initial viscosity under zero field, respectively. Kumar et al. reported a significant $\eta_{\%}$~700% at the N–$N_F$ phase transition in RM734 [35]. Our results closely match these findings ($\eta_{\%}$~800%) for RM734 (Figure 6d and Figures S2a and S2b). However, at the $M_{AF}$–$N_F$ phase transition for DIO, $\eta_{\%}$~7000% (Figure 6c), i.e., an order of magnitude larger than that at the N–$N_F$ phase transition for RM734. In RM734, during the N–$N_F$ phase transition, the ferroelectric domains develop rapidly resulting in a notable electroviscous effect [35]. Near the $M_{AF}$–$N_F$ phase transition of DIO, the viscosity exhibits critical behavior even under a zero $E$-field (Figure 6a, gray circles), likely owing to competing antiferroelectric and ferroelectric interactions. The application of an *E*-field in the $M_{AF}$ phase amplifies this effect, effectively frustrating and hardening the material as it transitions to the $N_F$ phase. This is consistent with the 2 °C shift of the transition temperature when an *E*-field is applied (Figure 6c). Recently, precise differential scanning calorimetry measurements have confirmed the presence of the $M_{AF}$ phase even in RM734 [36]. However, the dramatic electroviscous effect at the $M_{AF}$–$N_F$ phase transition is not observed in RM734. This difference may be attributable to various factors, specifically, the temperature range of the $M_{AF}$ phase is below 1 K, the molar transition enthalpy between $M_{AF}$ and $N_F$ is approximately half that of DIO, and the intermolecular interactions before and after the transition are weaker than those of DIO. The electro-rheological properties of the studied material can be described using the following phenomenological equation:

$$\eta = \eta_0(1 + AE^2), \qquad (3)$$

where $\eta_0$, *A*, and *E* denote the bulk viscosity, viscoelectric coefficient, and electric field, respectively. *A* can be defined as

$$\frac{\eta(E) - \eta_0}{\eta_0} = \frac{\Delta \eta}{\eta_0} = AE^2. \qquad (4)$$

Thus, *A* can be calculated from the slope of the relative viscosity $(\eta(E) - \eta_0)/\eta_0$ vs. $E^2$. For our system, *A* in the N, $M_{AF}$, and $N_F$ phases was estimated using Equation (4). The relative viscosity vs. $E^2$ curve is shown in Figures 5b and 5d. We derived the slope from the linear region. The estimated values were as follows: $A_N$ (95 °C) = 0.91 × $10^{-9}$, $A_{MAF}$ (74 °C) = 0.16 × $10^{-9}$, and $A_{NF}$ (60 °C) = 3.0 × $10^{-9}$ $V^2$ $m^{-2}$ (Figure 6b). For RM734, $A_N$ (150 °C) = 0.08 × $10^{-9}$ and $A_{NF}$ (110 °C) = 3.0 × $10^{-9}$ $V^2$ $m^{-2}$ under the same conditions (Figures S2c and S2d). Notably, the value of $A_{NF}$ for DIO and RM734 is 5–7 orders of magnitude greater than that exhibited by common organic polar fluids [37].

## 3. Conclusion

We investigated the rheological properties and electroviscous effects of the polar fluid DIO in the N, $M_{AF}$, and $N_F$ phases, corresponding to a P–AF–F system. At high shear rates, the $N_F$ phase exhibited an increase in effective viscosity (shear thickening) owing to the generation of polar defects. In the $M_{AF}$ phase, two types of shear-thinning modes were observed at low and high shear rates, and the apparent viscosity responses to increasing shear rate and shear stress were remarkable. The viscosity significantly decreased by this two-stage shear-thinning, which may be attributable to two potential mechanisms: the transition from the parallel arrangement of the $M_{AF}$ phase to the bookshelf arrangement, and the shift from flow orientation to tumbling behavior. Finally, we demonstrated that an ultra-low *E*-field (0.15 V μm$^{-1}$) induced an extremely large electroviscous effect (70-fold increase) near the modulated antiferroelectric–ferroelectric phase transition. The effect was an order of magnitude larger than that observed in RM734 near the paraelectric–ferroelectric phase transition. We believe that this unique electroviscous effect can be leveraged to develop an electrical smart brake suitable for smart cities and smart mobility solutions.

**4. Experimental section**

*4.1. Materials and general measurements*

The NF LC material (DIO) was synthesized and recrystallized twice from dichloromethane/n-hexane mixture in our laboratory. The NF LC material (RM734) was purchased from INSTEC Inc. and used without any purifications. The rheological properties of the NF LCs were measured by an Anton Paar MCR502 rheometer by using a parallel plate measuring system (PP25/DI/TI, diameter ~25 mm and PP50/DI/TI-SN38600, diameter ca. 50 mm), as shown in Figure S3a. The sample thickness was fixed to be 80 μm for all experiments. All measurements were conducted on cooling from the isotropic phase under N2 gas stream to reduce sample degradation.

*4.2. Measurement system #1*

The rheometer was equipped with the Anton Paar P-PTD 200/SS/DI dielectric module to investigate the electro-viscous effect in LC phases. We turned on potential difference between the fixed bottom and the rotating measuring system by using the signals of a TiePie Handyscope HS3 device amplified by an FLC F10A amplifier. The electrical contact to the rotating part was achieved by a needle orbiting in a 3D printed circular pool filled with 10% aqueous NaCl solution to eliminate friction. To avoid water evaporation from the electrolyte we applied a covering layer of silicone oil (Wacker Silicone Fluids AK M20, viscosity: 20 mPa·s@r.t.). We confirmed that the torque from the fluid electrical contact was too small to affect any measurement.

*4.3. Measurement system #2*

To observe polarized optical microscopic (POM) textures without and with *E*-field, we used the Anton Paar rheo-microscope module with a custom-made parallel plate measuring system made of an indium-tin-oxide (ITO) coated glass (diameter: ca. 25 mm), and a custom-made monochromatic LED light source (wavelength: 660 nm) for transmission mode (Figure S3b).

**Acknowledgement**

This work was financially supported by the Hungarian National Research, Development, and Innovation Office under grants NKFIH FK142643 and 2023-1.2.1-ERA_NET-2023-00008 (P.S.), the US National Science Foundation under grant DMR-2210083 (A.J.). The work was also supported by the János Bolyai Research Scholarship of the Hungarian Academy of Sciences (HAS) (BO/00294/22/11; P.S.). This work was partially supported by JSPS KAKENHI (JP22K14594; H.N., 23K17341, JP21H01801; F.A.), RIKEN Special Postdoctoral Researchers (SPDR) fellowship (H.N.), FY2022 RIKEN Incentive Research Projects (H.N.), and JST CREST (JPMJCR17N1; F.A.) and JST SICORP EIG CONCERT-Japan (JPMJSC22C3; F.A.).

**References**

[1] Cummins, H. Z.: Experimental studies of structurally incommensurate crystal phases, *Phys. Rep.* **1990**, *185*, 211, DOI: 10.1016/0370-1573(90)90058-A

[2] Blinc, R.; Levanyuk, A. P.: In *Incommensurate Phases in Dielectrics*, 1st ed..; North-Holland, 1986.

[3] Wilson, J. A.; Disalvo F. J.; Mahajan, S.: Charge-density waves and superlattices in the metallic layered transition metal dichalcogenides, *Adv. Phys.* **1975**, *24*, 117–120, DOI: 10.1080/00018737500101391

[4] Rossat-Mignod, J.: Magnetic structures of rare earth intermetallics, *J. Phys. Colloquies*, **1979**, *40*, C5-95–C5-100, DOI: 10.1051/jphyscol:1979535

[5] Aruga, T.: Charge-density waves on metal surfaces, *J. Phys.: Condens. Matter* **2002**, *14*, 8393, DOI: 10.1088/0953-8984/14/35/310

[6] Clarke, R.; Caswell N.; Solin, S. A.; Horn, P. M.: Positional and orientational correlations in the graphite intercalate $C_{24}Cs$, *Phys. Rev. Lett.* **1979**, *43*, 2018, DOI: doi.org/10.1103/PhysRevLett.43.2018

[7] Yamamoto, A.: Modulated structure of wustite ($Fe_{1-x}O$) (three-dimensional modulation), *Acta Cryst*. **1982**, *B38*(5), 1451–1456, DOI: 10.1107/S056774088200613X

[8] Cady, A.; Han, X. F.; Olson, D. A.; Orihara H.; Huang, C. C.: Optical characterization of a nanoscale incommensurate pitch in a new liquid-crystal phase, *Phys. Rev. Lett.*, **2003**, *91*, 125502, DOI: 10.1103/PhysRevLett.91.125502

[9] Onoda M.; Kato, K.: Structure of the incommensurate composite crystal $(PbS)_{1.12}VS_2$, *Acta Cryst*. **1990**, *B46*, 487–492, DOI: 10.1107/S0108768190003950

[10] Shaw, T. M.; Shivashankar, S. A.; La Placa, S. J.; Cuomo, J. J.; McGuire, T. R.; Roy, R. A.; Kelleher K. H.; Yee, D. S.: Incommensurate structure in the Bi-Sr-Ca-Cu-O 80-K superconductor, *Phys. Rev. B*, **1988**, *37*, 9856, DOI: 10.1103/PhysRevB.37.9856

[11] Tanisaki, S.: Microdomain structure in paraelectric phase of $NaNO_2$, *J. Phys. Soc. Jpn.*, **1961**, *16*, 579, DOI: 10.1143/JPSJ.16.579

[12] Shiozaki, Y.: Satellite X-ray scattering and structural modulation of thiourea, *Ferroelectrics*, **1971**, *2*, 245–260, DOI: 10.1080/00150197108234099

[13] a) Nishikawa, H.; Shiroshita, K.; Higuchi, H.; Okumura, Y.; Haseba, Y.; Yamamoto, S.; Sago, K.; Kikuchi, H.: A fluid liquid-crystal material with highly polar order, *Adv. Mater*. **2017**, *29*, 1702354, DOI: 10.1002/adma.201702354; b) Mandle, R. J.; Cowling, S. J.; Goodby, J. W.: A nematic to nematic transformation exhibited by a rod-like liquid crystal, *Phys. Chem. Chem. Phys*. **2017**, *19*, 11429–11435, DOI: 10.1039/C7CP00456G; c) Chen, X.; Korblova, E.; Dong, D.; Wei, X.; Shao, R.; Radzihovsky, L.; Glaser, M. A.; MacLennan, J. E.; Bedrov, D.; Walba, D. M.; Clark, N. A.: First-principles experimental demonstration of ferroelectricity in a thermotropic nematic liquid crystal: Polar domains and striking electro-optics, *Proc. Natl. Acad. Sci. USA* **2020**, *117*(25), 14021–14031, DOI: 10.1073/pnas.2002290117; d) Manabe, A.; Bremer, M.; Kraska, M.: Ferroelectric nematic phase at and below room temperature, *Liq. Cryst*. **2021**, *48*, 1079–1086, DOI: 10.1080/02678292.2021.1921867; e) Sebastián, N.; Čopič, M. ; Mertelj, A.: Ferroelectric nematic liquid-crystalline phases, *Phys. Rev. E*. **2022**, *106*, 021001, DOI: 10.1103/PhysRevE.106.021001

[14] a) Mertelj, A.; Cmok, L.; Sebastián, N.; Mandle, R. J.; Parker, R. R.; Whitwood, A. C.; Goodby, J. W.; Čopič, M.: Splay nematic phase, *Phys. Rev. X* **2018**, *8*, 041025, DOI: Phys. Rev. X 2018, 8, 041025; b) Sebastián, N.; Cmok, L.; Mandle, R. J.; De La Fuente, M. R.; Drevenšek Olenik, I.; Čopič, M. ; Mertelj, A.: Ferroelectric-ferroelastic phase transition in a

nematic liquid crystal, *Phys. Rev. Lett.* **2020**, *124*, 037801, DOI: Phys. Rev. Lett. 2020, 124, 037801

[15] a) Basnet, B.; Rajabi, M.; Wang, H.; Kumari, P.; Thapa, K.; Paul, S.; Lavrentovich, M. O.; Lavrentovich, O. D.: Soliton walls paired by polar surface interactions in a ferroelectric nematic liquid crystal, *Nat. Commun.* **2022**, *13*, 3932, DOI: 10.1038/s41467-022-31593-w; b) Kumari, P.; Basnet, B.; Wang, H.; Lavrentovich, O. D.: Ferroelectric nematic liquids with conics, *Nat. Commun.* **2023**, *14*, 748, DOI: 10.1038/s41467-023-36326-1; c) Sebastián, N.; Lovšin, M.; Berteloot, B.; Osterman, N.; Petelin, A.; Mandle, R. J.; Aya, S.; Huang, M.; Drevenšek-Olenik, I.; Neyts, K.; Mertelj, A.: Polarization patterning in ferroelectric nematic liquids via flexoelectric coupling, *Nat. Commun.* **2023**, *14*, 3029, DOI: 10.1038/s41467-023-38749-2; d) Yang, J.; Zou, Y.; Tang, W.; Li, J.; Huang, M.; Aya, S.: Spontaneous electric-polarization topology in confined ferroelectric nematics, *Nat. Commun.*, **2023**, *13*, 7806, DOI: 10.1038/s41467-022-35443-7; e) Kumari, P., Basnet, B., Lavrentovich, M. O., Lavrentovich, O. D.: Chiral ground states of ferroelectric liquid crystals, *Science* **2024**, *83*, 1364–1368, DOI: 10.1126/science.adl0834; f) Yang, J.; Zou, Y.; Huang, M.; Aya. S.: Flexoelectricity-driven toroidal polar topology in liquid-matter helielectrics, *Nat. Phys.* **2024**, *20*, 991–1000, DOI: 10.1038/s41567-024-02439-7

[16] Máthé, M. T.; Buka, Á.; Jákli, A.; Salamon, P.: Ferroelectric nematic liquid crystal thermomotor, *Phys. Rev. E.* **2022**, *105*, L052701, DOI: 10.1103/PhysRevE.105.L052701

[17] a) Marni, S., Nava, G.; Barboza, R.; Bellini, T. G., Lucchetti, L.: Walking ferroelectric liquid droplets with light, *Adv. Mater.*, **2023**, *35*, 2212067, DOI: 10.1002/adma.202212067; b) Máthé, M. T., Nishikawa, H., Araoka, F., Jákli, A., Salamon, P.: Electrically activated ferroelectric nematic microrobots, *Nat. Commun.* **2024**, *15*, 6928, DOI: 10.1038/s41467-024-50226-y

[18] a) Máthé, M. T.; Perera, K.; Buka, Á.; Salamon, P.; Jákli, A.: Fluid ferroelectric filaments, *Adv. Sci.* **2024**, *11*, 2305950, DOI: 10.1002/advs.202305950; b) Jarosika, A., Nádasi, H., Schwidder, M., Manabe, A., Bremer, M., Memmer, M. K.; Eremin, A.: Fluid fibers in true 3D ferroelectric liquids, *Proc. Natl. Acad. Sci. USA* **2024**, *121*(13), e2313629121, DOI: 10.1073/pnas.2313629121

[19] a) Barboza, R.; Marni, S.; Ciciulla, F.; Mir, F. A.; Nava, G.; Caimi, F.; Bellini, T.; Lucchetti, L.: Explosive electrostatic instability of ferroelectric liquid droplets on ferroelectric solid surfaces, *Proc. Natl. Acad. Sci. USA* **2022**, *119*(32), e2207858119, DOI: 10.1073/pnas.2207858119; b) Máthé, M. T.; Farkas, B.; Péter, L.; Buka, Á.; Jákli, A.; Salamon, P: Electric field-induced interfacial instability in a ferroelectric nematic liquid crystal, *Sci. Rep.* **2023**, *13*(1), 6981, DOI: 10.1038/s41598-023-34067-1; c) Marni, S.; Caimi, F.; Barboza, R.; Clark, N.; Bellini, T.; Lucchetti, L.: Fluid jets and polar domains, on the relationship between electromechanical instability and topology in ferroelectric nematic liquid crystal droplets, *Soft Matter*, **2024**, *20*(25), 4878–4885, DOI: 10.1039/d4sm00317a

[20] Caimi, F.; Nava, G.; Fuschetto, S.; Lucchetti, L.; Paiè, P.; Osellame, R.; Chen, X.; Clark, N. A.; Glaser, M. A.; Bellini, T.: Fluid superscreening and polarization following in confined ferroelectric nematics, *Nat. Phys.* **2023**, *19*, 1658–1666, DOI: 10.1038/s41567-023-02150-z

[21] a) Nishikawa, H.; Araoka, F.: A new class of chiral nematic phase with helical polar order, *Adv. Mater.* **2021**, 33, 2101305, DOI: 10.1002/adma.202101305, b) Zhao, X.; Zhou, J.; Li., J.; Kougo, J.; Wan, Z.; Huang. M.; Aya. S.: Spontaneous helielectric nematic liquid crystals: Electric analog to helimagnets, *Proc. Natl. Acad. Sci. U.S.A.* **2021**, *118*(42): e2111101118, DOI: 10.1073/pnas.2111101118; c) Feng, C.; Saha, R.; Korblova, E.; Walba, D.; Sprunt, S. N.; Jákli, A.: Electrically tunable reflection color of chiral ferroelectric nematic liquid crystals, *Adv. Opt. Mater.* **2021**, *9*, 2101230, DOI: 10.1002/adom.202101230; d) Karcz, J.; Herman, J.; Rychłowicz, N.; Kula, P.; Górecka, E.; Szydlowska, J.; Majewski, P. W.; Pociecha, D.: Spontaneous symmetry breaking in polar fluids, *Science* **2024**, *384*, 1096–1099, DOI: 10.1038/s41467-024-50230-2; e) Gibb, C. J., Hobbs, J., Nikolova, D. I., Raistrick, T., Berrow, S. R., ertelj, A., Osterman, N., Sebastián, N., Gleeson, H. F., Mandle, R. J.: Spontaneous symmetry breaking in polar fluids, *Nat. Commun.*, **2024**, *15*, 5845, DOI: 10.1038/s41467-024-50230-2; f) Nishikawa, H.; Okada, D.; Kwaria, D.; Nihonyanaghi, A.; Kuwayama, M.; Hoshino, M.; Araoka, F.: Emergent ferroelectric nematic and heliconical ferroelectric nematic states in an achiral "straight" polar rod mesogen, *Adv. Sci.* **2024**, *11*, 2405718, DOI: 10.1002/advs.202405718

[22] Chen, X.; Martinez, V.; Korblova, E.; Freychet, G.; Zhernenkov, M.; Glaser, M. A.; Wang, C.; Zhu, C.; Radzihovsky, L.; Maclennan, J. E.; Walba, D. M.; Clark, N. A.: The smectic ZA

phase: Antiferroelectric smectic order as a prelude to the ferroelectric nematic., *Proc. Natl. Acad. Sci. USA* **2023**, *120*(8), e2217150120, DOI: 10.1073/pnas.2217150120

[23] a) Qiu, S. L.; Dutta, M.; Cummins, H. Z.; Wicksted, J. P.; Shapiro, S. M.; Extension of the Lifshitz-point concept to first-order phase transitions: incommensurate $NaNO_2$ in a transverse electric field, *Phys. Rev. B.*, **1986**, *34*(11), 7901–7910, DOI: 10.1103/PhysRevB.34.7901; b) Scott, J. F.: Prospects for Ferroelectrics: 2012–2022. *Int. Sch. Res. Notices*, **2013**, 1–24, DOI: 10.1155/2013/187313

[24] a) Jamet, J. P.: Electric field phase diagram of thiourea determined by optical birefringence. *J. Phys. Lett. (France)*, **1981**, *42*, 123–125, DOI: 10.1051/jphyslet:01981004206012300; b) Lederer, P.; Chaves, C. M.: Phase diagram of thiourea at atmospheric pressure under electric field: a theoretical analysis. *J. Phys. Lett. (France)*, **1981**, *42*, 127–130, DOI: 10.1051/jphyslet:01981004206012700

[25] Cruickshank, E.; Rybak, P.; Majewska, M. M.; Ramsay, S.; Wang, C.; Zhu, C.; Walker, R.; Storey, J. M. D.; Imrie, C. T.; Gorecka, E.; Pociecha. D.: To be or not to be polar: the ferroelectric and antiferroelectric nematic phases, *ACS Omega* **2023**, *8*, 36562–36568, DOI: 10.1021/acsomega.3c05884

[26] Nacke, P.; Manabe, A.; K.–Memmer, M.; Chen, X.; Martinez, V.; Freychet, G.; Zhernenkov, M.; Maclennan, J. E.; Clark, N. A.; Bremer, M.; Giesselmann, F.: New examples of ferroelectric nematic materials showing evidence for the antiferroelectric smectic-Z phase, *Sci. Rep.* **2024**, *14*(1), 4473, DOI: 10.1038/s41598-024-54832-0

[27] de Gennes, P. G.; Prost. J.: The physics of liquid crystals, 2nd ed., Oxford University Press, New York, 1993.

[28] Fisher, J.; Frederickson, A. G.: Transport processes in anisotropic fluids II. coupling of momentum and energy transport in a nematic mesophase, *Mol. Cryst. Liq. Cryst.*, **1969**, *6*(2), 255–271, DOI: 10.1080/15421406908082963

[29] Spriggs, T. W.; Huppler, J. D.; Bird, R. B.: An experimental appraisal of viscoelastic models, *Trans. Soc. Rheol.*, **1966**, 10, 191–213, DOI: 10.1122/1.549057

[30] Erickesen, J. L.: A Boundary-layer effect in viscometry of liquid crystals, *Trans. Soc. Rheol*, **1969**, *13*, 9–15, DOI: 10.1122/1.549158

[31] a) Mather, P. T.; Pearson, D. S.; Larson, R. G.: Flow patterns and disclination-density measurements in sheared nematic liquid crystals I: Flow-aligning 5CB, *Liq. Cryst.*, **1996**, *20*(5), 527–538, DOI: 10.1080/02678299608031139; b) Mather, P. T.; Pearson, D. S.; Larson, R. G.: Flow patterns and disclination-density measurements in sheared nematic liquid crystals II: Tumbling 8CB, *Liq. Cryst.*, **1996**, *20*(5), 539–546, DOI: 10.1080/02678299608031140

[32] Stewart, I. W. In *The Static and Dynamic Continuum Theory of Liquid Crystals: A Mathematical Introduction*, 1st ed., CRC Press, 2004. DOI: 10.1201/9781315272580

[33] a) Negita, K.; Uchino, S.: Rheological study on the shear-induced structural changes in liquid crystalline phases of octylcyanobiphenyl, *Mol. Cryst. Liq. Cryst.*, **2002**, *378*(1), 103–112, DOI: 10.1080/713738584; b) Negita, K.; Inoue, M.; Kondo, S.: Dielectric study on the shear-induced structural changes in the nematic and the smectic-*A* phases of 4-n-octyl-4'-cyanobiphenyl (8CB), *Phys. Rev. E*, **2006**, *74*, 051708, DOI: 10.1103/PhysRevE.74.051708; c) Yamamoto, T.; Nagae, Y.; Wakabayashi, T.; Kamiyama, T.; Suzuki, H.: Calorimetry of phase transitions in liquid crystal 8CB under shear flow, *Soft Matter*, **2023**, *19*, 1492–1498, DOI: 10.1039/d2sm01652d

[34] a) Szydlowska, J.; Majewski, P.; Čepič, M.; Vaupotič, N.; Rybak, P.; Imrie, C. T.; Walker, R.; Cruickshank, E.; Storey, J. M. D.; Damian, P.; Gorecka, E.: Ferroelectric nematic-isotropic liquid critical end point, *Phys. Rev. Lett.* **2023**, *130*(21), 216802, DOI: 10.1103/PhysRevLett.130.216802; b) Mrukiewicz, M.; Perkowski, P.; Karcz J.; Kula, P.: Ferroelectricity in a nematic liquid crystal under a direct current electric field, *Phys. Chem. Chem. Phys.*, **2023**, *25*, 13061, DOI: 10.1039/D3CP00714F; c) Adaka, A.; Guragain, P.; Perera, K.; Nepal, P.; Twieg, R. J.; Jákli, A.: Low field electrocaloric effect at isotropic–ferroelectric nematic liquid transition, *Soft Matter*, **2025**, 21, 458–462, DOI: 10.1039/D4SM00979G

[35] Kumar, M. P.; Karcz, J.; Kula, P.; Karmakar, S.; Dhara, S.: Giant electroviscous effects in a ferroelectric nematic liquid crystal, *Phys. Rev. A*, **2023**, *19*, 044082, DOI: 10.1103/PhysRevApplied.19.044082

[36] a) Thoen, J.; Cordoyiannis, G.; Korblova, E.; Walba, D. M.; Clark, N. A.; Jiang, W.; Mehl, G. H.; Glorieux, C.: Calorimetric evidence for the existence of an intermediate phase between

the ferroelectric nematic phase and the nematic phase in the liquid crystal RM734, *Phys. Rev. E*, **2024**, *110*, 014703, DOI: 10.1103/PhysRevE.110.014703; b) Thoen, J.; Cordoyiannis, G.; Jiang, W.; Mehl, G. H.; Glorieux, C.: Phase transitions study of the liquid crystal DIO with a ferroelectric nematic, a nematic, and an intermediate phase and of mixtures with the ferroelectric nematic compound RM734 by adiabatic scanning calorimetry, *Phys. Rev. E*, **2023**, *107*, 014701

[37] a)Hunter, R. J.; Leyendekkers, J. V.: Viscoelectric coefficient for water, *J. Chem. Soc., Faraday Trans*. 1, **1978**, *74*, 450, DOI: 10.1039/F19787400450; b) Jin, D.; Hwang, Y.; Chai, L.; Kampf, N.; Klein, J.: Direct measurement of the viscoelectric effect in water, *Proc. Natl. Acd. Sci. USA* **2022**, *119*(1), e2113690119, DOI: 10.1073/pnas.2113690119

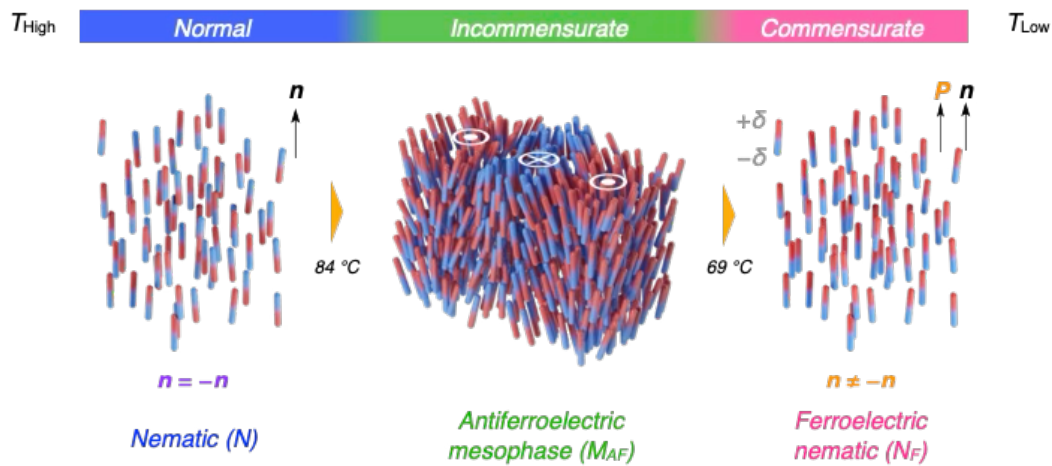

**Figure 1**. Illustration of the normal (N) – incommensurate ($M_{AF}$)–commensurate ($N_F$) phase sequence in DIO. **n** and **P** denote orientation polarization directors, respectively. $+\delta$ and $-\delta$ denote plus and minus electrical charges at one end of a molecule, respectively.

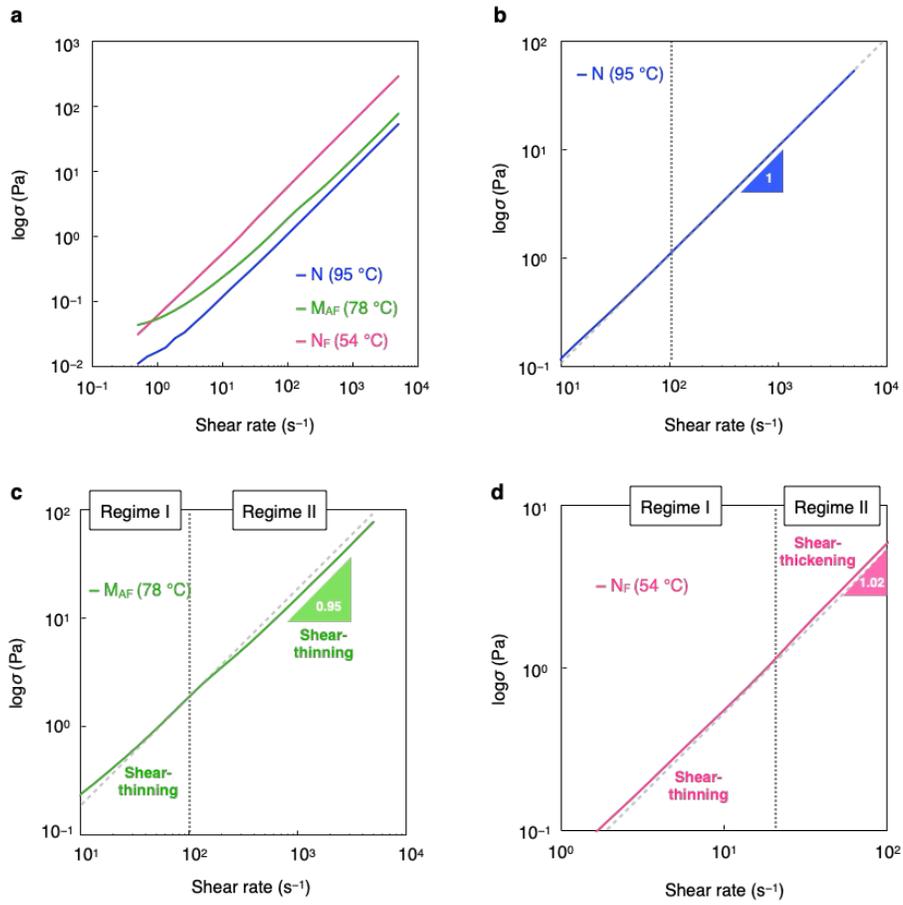

**Figure 2**. a) Stress as a function of shear rate in the N, $M_{AF}$ and $N_F$ phases. Enlarged panel for N (b), $M_{AF}$ (c) and $N_F$ (d) phases.

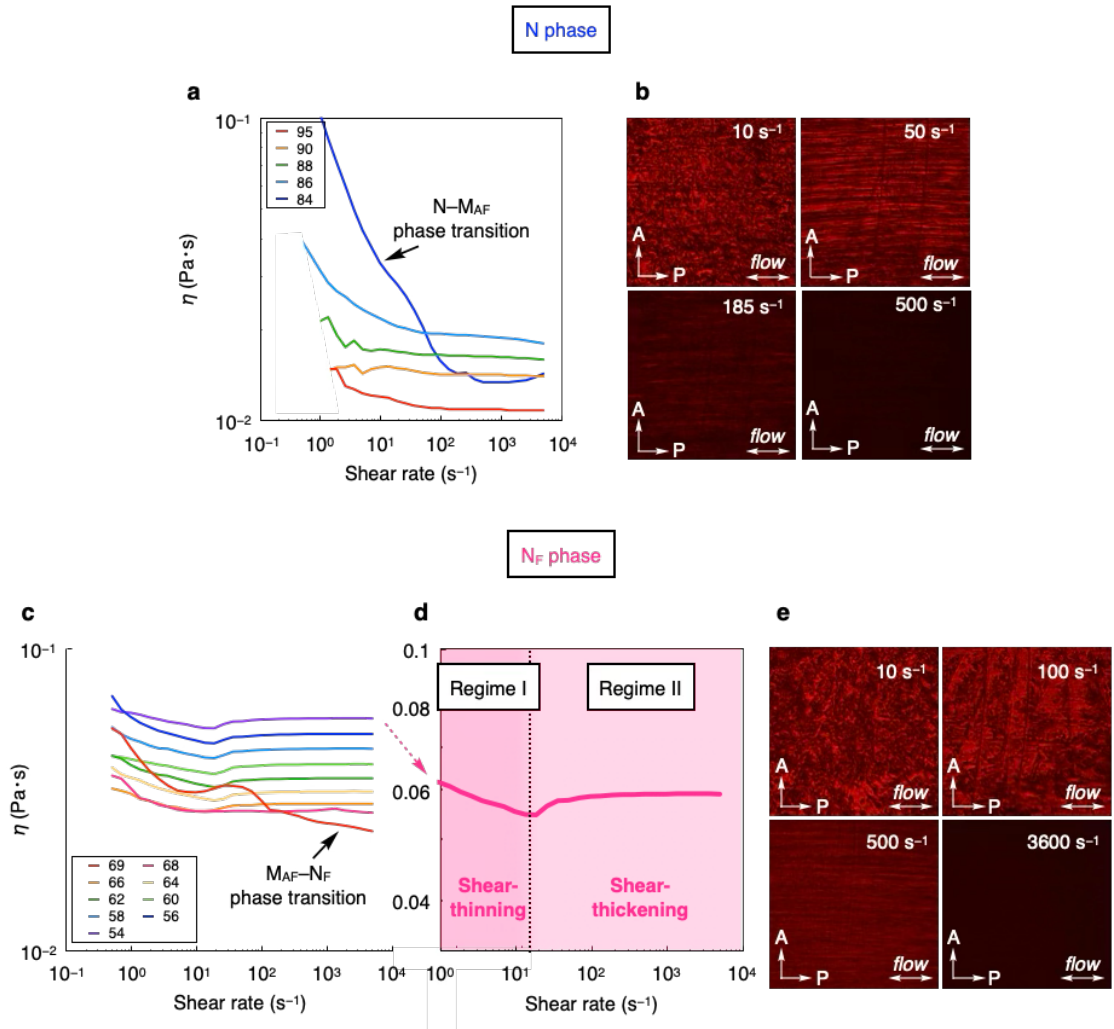

**Figure 3**. Apparent viscosity as a function of shear rate and the corresponding POM images during shearing with different shear rate in the N (a,b) and $N_F$ (c,d,e) phases. The POM images were taken at 95 °C (b) and 60 °C (e). (d) Highlighted profile recorded at 54 °C.

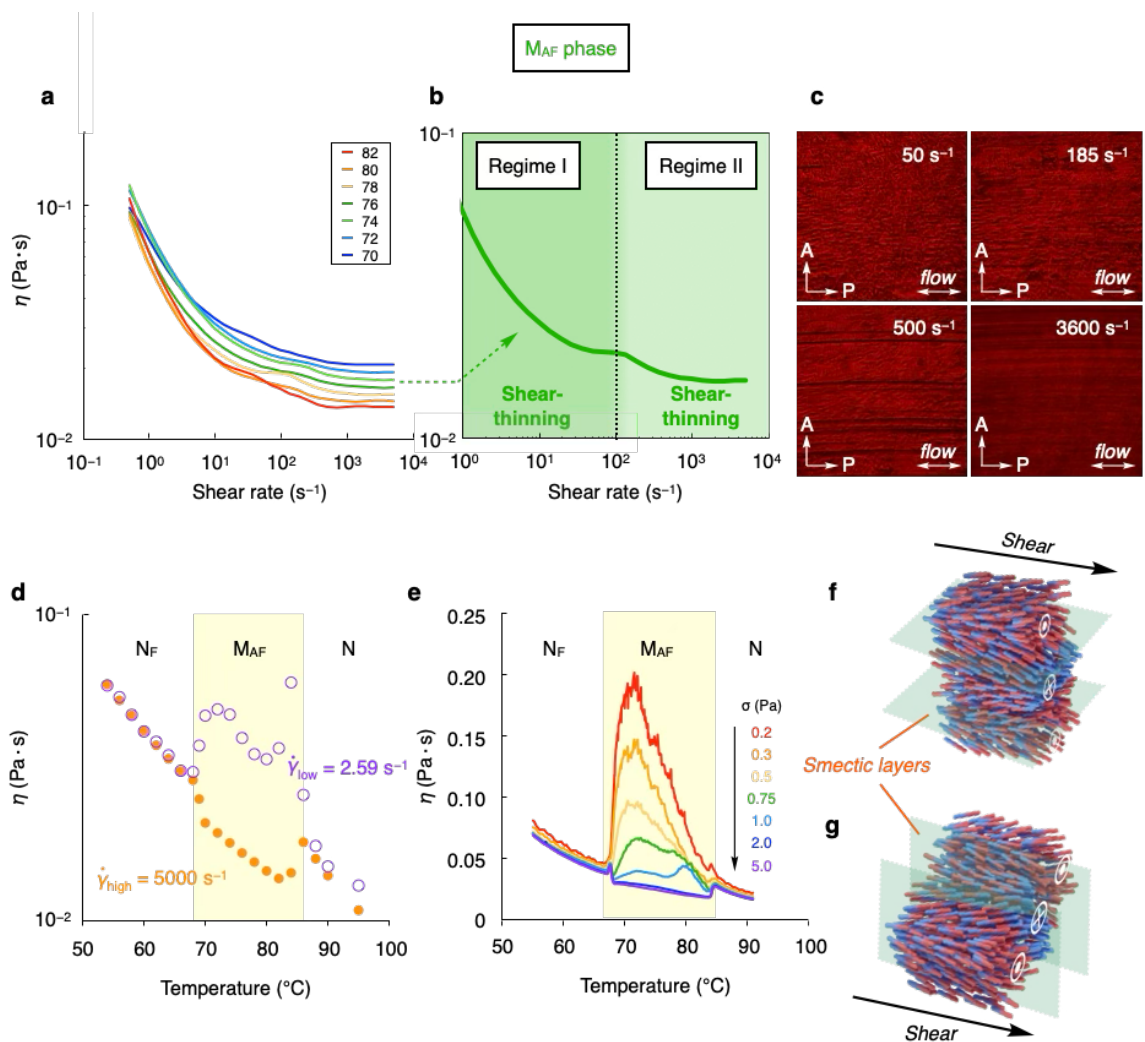

*Figure 4*. (a) Apparent viscosity as a function of shear rate at different temperatures. (b) Highlighted profile recorded at 78 °C; (c) The corresponding POM images taken at 78 °C during shearing with different shear rate; (d) Apparent viscosity vs temperature recorded at high (5000 s$^{-1}$) and low (2.59 s$^{-1}$) shear rate; (e) Apparent viscosity vs temperature recorded at various shear stress ranging between 0.2 and 5.0 Pa. Schematic illustration of M$_{AF}$ phase (parallel (f) and bookshelf (g) structure*s*), in which each molecule lies on the substrate and flow along shear direction. In the parallel and bookshelf structures, the smectic layer is parallel and perpendicular to the substrate, respectively.

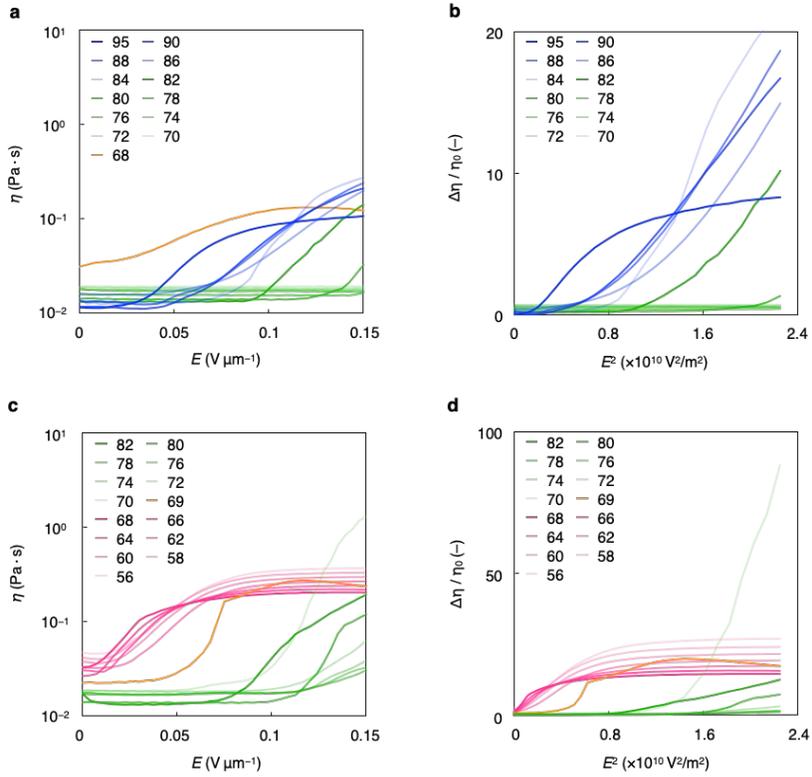

**Figure 5**. Electric field dependence of the apparent viscosity measured at constant shear rate (100 s$^{-1}$) by applying (a):1 kHz $E_{AC}$-field; (c): $E_{DC}$-field in the N (blue line), M$_{AF}$ (green line), and N$_F$ (magenta line) phases. (b,d): $\Delta\eta/\eta_0$ vs $E^2$ profiles obtained from data in the panel (a) and (c), respectively. Note that during the M$_{AF}$–N$_F$ phase transition, the light transmittance was almost zero because the director reoriented along the $E$-field.

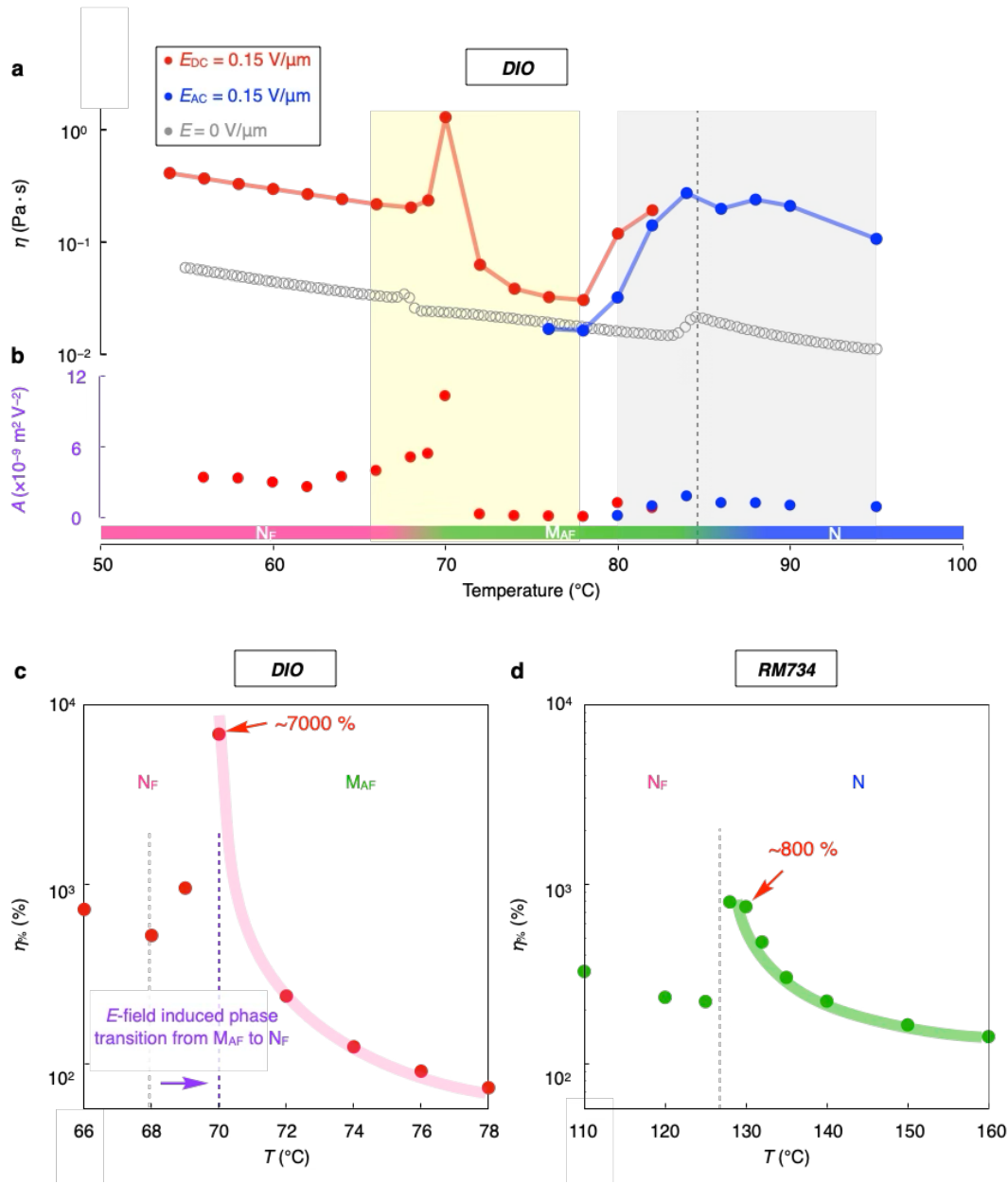

**Figure 6**. (a): Temperature dependence of the apparent viscosity under constant ($\sigma = 20$ Pa) shear stress without/with $E$-field ($E_{DC} = E_{AC} = 0.15$ V μm$^{-1}$, $f(E_{AC}) = 1$ kHz); (b): Visco-electric constant A as a function of temperature; (c): $\eta_\%$ vs T for DIO; (c): $\eta_\%$ vs T for RM734 (d).

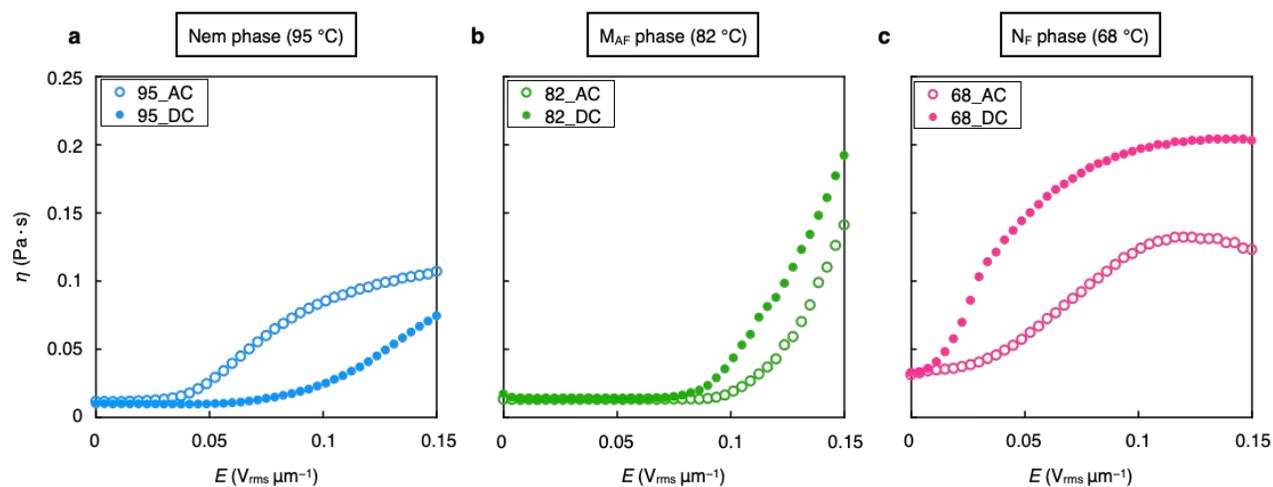

**Figure S1** Apparent viscosity as a function of $E$-field for N (a), $M_{AF}$ (b) and $N_F$ (c) phases. Open and closed circle plots denote data recorded under AC and DC $E$-field, respectively.

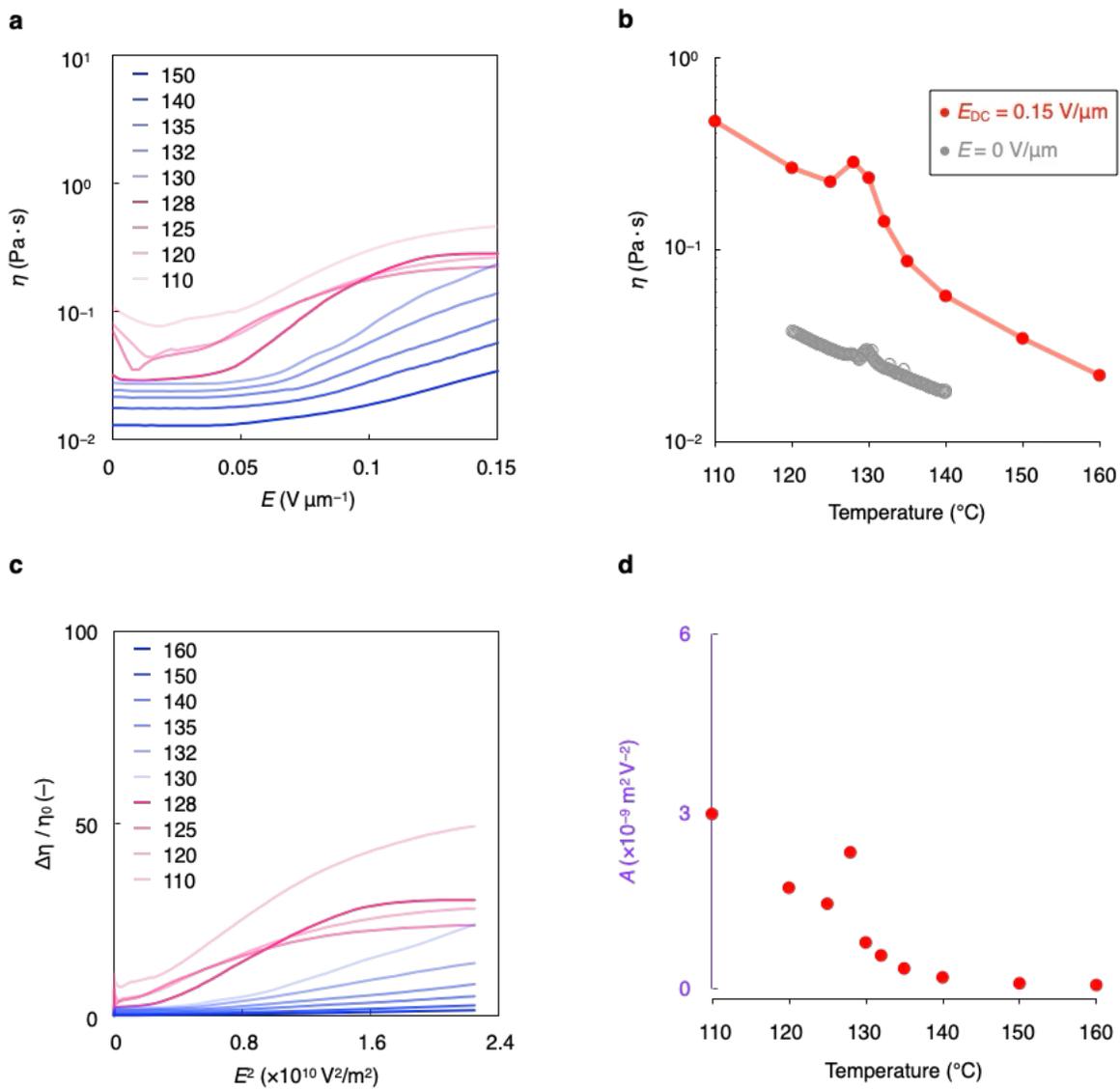

**Figure S2** Electro-rheological properties of RM734. a) Apparent viscosity evolution under 120 s$^{-1}$ shear rate by applying $E_{DC}$-field in the N (blue line) and N$_F$ (magenta line) phases. b) Temperature dependence of the apparent viscosity under $\sigma$ = 20 Pa shear stress without/with $E$-field ($E_{DC} = E_{AC}$ = 0.15 Vrms μm$^{-1}$, $f(E_{AC})$ = 1 kHz). c) Δη/η$_0$ vs $E^2$ profiles obtained from data in the panel (a). d) Viscoelectric constant $A$ as a function of temperature.

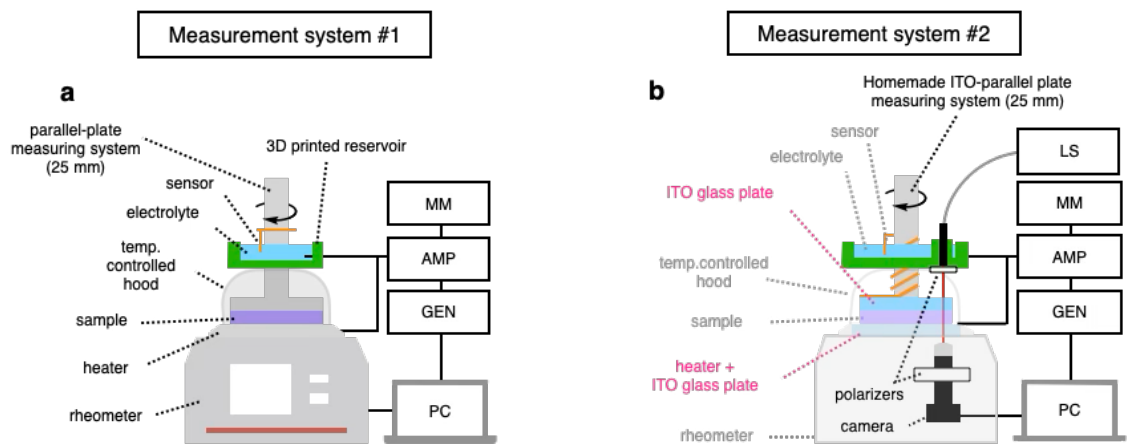

**Figure S3** Schematic illustration of measurement setups. We used 10% aqueous NaCl solution as an electrolyte. To prevent its evaporation, we covered the bottom aqueous layer with upper silicone oil layer. Abbreviations: MM = multimeter, AMP = amplifier, GEN = function generator and LS = monochromatic LED light source (wavelength: 660 nm).